\documentclass[12pt]{iopart}
\usepackage{graphicx}
\usepackage{xcolor}

\usepackage{siunitx}
\usepackage{ulem}
\expandafter\let\csname equation*\endcsname\relax
\expandafter\let\csname endequation*\endcsname\relax
\usepackage{amsmath}
\usepackage{amssymb}
\usepackage[colorlinks=False]{hyperref}
\begin{document}
\title[Preparation of one $^{87}$Rb and one $^{133}$Cs atom in a single optical tweezer]{Preparation of one $^{87}$Rb and one $^{133}$Cs atom in a single optical tweezer}

\author{R V Brooks$^1$, S Spence$^1$, A Guttridge$^1$, A Alampounti$^1$, A Rakonjac$^1$, L A McArd$^1$, Jeremy M Hutson$^2$ and Simon L Cornish$^1$}

\address{$^1$ Joint Quantum Centre (JQC) Durham-Newcastle, Department of Physics, Durham University, South Road, Durham DH1 3LE, United Kingdom

$^2$ Joint Quantum Centre (JQC) Durham-Newcastle, Department of Chemistry, Durham University, South Road, Durham DH1 3LE, United Kingdom}

\begin{abstract}
\noindent
We report the preparation of exactly one $^{87}$Rb atom and one $^{133}$Cs atom in the same optical tweezer as the essential first step towards the construction of a tweezer array of individually trapped $^{87}$Rb$^{133}$Cs molecules. Through careful selection of the tweezer wavelengths, we show how to engineer species-selective trapping potentials suitable for high-fidelity preparation of Rb $+$ Cs atom pairs. Using a wavelength of 814~nm to trap Rb and 938~nm to trap Cs, we achieve loading probabilities of 0.508(6) for Rb and 0.547(6) for Cs using standard red-detuned molasses cooling. Loading the traps sequentially yields exactly one Rb and one Cs atom in $28.4(6)\,\%$ of experimental runs. Using a combination of an acousto-optic deflector and a piezo-controlled mirror to control the relative position of the tweezers, we merge the two tweezers, retaining the atom pair with a probability of $0.99^{(+0.01)}_{(-0.02)}$. We use this capability to study hyperfine-state-dependent collisions of Rb and Cs in the combined tweezer and compare the measured two-body loss rates with coupled-channel quantum scattering calculations.

\end{abstract}


\section{Introduction}

Optical tweezers have emerged as a powerful experimental technique for quantum science owing to the inherent capability to prepare, address and detect single neutral atoms. 
Utilising optical tweezers, it is now possible to produce large filled arrays of single atoms using dynamic rearrangement \cite{Lee2016, Barredo2016, Endres2016, Stuart2018, OhlDeMello2019} and enhanced loading techniques \cite{Grunzweig2010, Brown2019}. Local and global control of the internal atomic degrees of freedom has been demonstrated using optical and microwave transfer techniques \cite{Schrader2004, Jones2007, Xia2015} and complete control over the motional degrees of freedom has been demonstrated using Raman sideband cooling (RSC) \cite{Kaufman2012,Thompson2013, Sompet2017, Liu2019, Wang2019} to transfer atoms to the motional ground state of the tweezer. The pristine, well-controlled environment achievable using optical tweezers has allowed studies of ultracold atomic collisions where the number of participants in a collision is exactly known \cite{Fuhrmanek2012, Xu2015, Sompet2019, Reynolds2020}. In addition, optical tweezers have led to significant progress in the fields of quantum simulation \cite{Browaeys2020} and quantum information processing \cite{Saffman2016,Levine2019} with individually controlled neutral atoms, where excitation to highly excited Rydberg states is utilised to engineer long-range interactions between the particles. In recent years, tweezer control has been extended beyond alkali-metal atoms to alkaline earth and alkaline-earth-like atoms \cite{Cooper2018,Norcia2018, Jackson2019, Saskin2019} further enriching the systems and tools available to experimentalists.

%
%
Over a similar time frame, the production and control of ultracold polar molecules has progressed enormously, with many groups now capable of producing molecules in their rovibrational ground state at ultracold temperatures, either by association of ultracold atoms \cite{Ni2008, Takekoshi2014, Molony2014, Park2015, Guo2016, Rvachov2017, Seeselberg2018a,  Yang2019, Hu2019, Voges2020} or by direct laser cooling \cite{Barry2014, Truppe2017, Anderegg2018, Collopy2018}. The strong drive to produce and control polar molecules is motivated by their intrinsic long-ranged dipolar interaction \cite{Carr2009, Yan2013} and their rich internal structure that provides many long-lived rotational and hyperfine states. These properties offer opportunities for studies of chemistry in the quantal regime \cite{Krems2008,deMiranda2011}, as well as applications in quantum simulation \cite{Barnett2006, Micheli2006, Gorshkov2011, Blackmore2019} and quantum computation \cite{DeMille2002, Yelin2006}. 







A natural development is to apply the control provided by optical tweezers to ultracold molecules. Confining individual molecules in arrays of optical tweezers will generate new possibilities for the implementation of quantum gates with polar molecules \cite{Ni2018, Sawant2020, Hughes2020} and for the simulation of gauge theories \cite{Luo2020}. Recent efforts in this direction have been met with success using both laser-cooled molecules \cite{Anderegg2019,Cheuk2020} and molecules formed using association techniques \cite{Liu2018, He2020,Zhang2020,Yu2020,Cairncross2021}. Ultimately, it is desirable to prepare the molecule in the motional ground state of the tweezer. Cooling a molecule to the motional ground state using Raman sideband cooling is more challenging than for atoms, though may be feasible \cite{Caldwell2020}. In contrast, molecules formed by association may be prepared in the motional ground state by first preparing an atom pair in the motional ground state of the tweezer \cite{Zhang2020, Cairncross2021}. In this approach, the constituent atoms are first loaded into separate optical tweezers and cooled to their motional ground states using RSC \cite{Kaufman2012,Thompson2013, Sompet2017, Liu2019, Wang2019}. The atoms are then merged into a single optical tweezer and subsequently converted into a ground-state molecule using a combination of magnetoassociation and STImulated Raman Adiabatic Passage (STIRAP) \cite{Bergmann1998} following established techniques. Crucially, molecules formed in this manner inherit the high occupancy of the motional ground state of the tweezer from the pre-cooled atoms.  Such a scheme has recently been used to produce a single $^{23}$Na$^{133}$Cs molecule in the rovibrational ground state \cite{Zhang2020, Cairncross2021}. We aim to use this scheme to produce arrays of single $^{87}$Rb$^{133}$Cs molecules in optical tweezers. For this molecule, both the magnetoassociation \cite{Takekoshi2012, Koppinger2014} and STIRAP \cite{Takekoshi2014, Molony2014} steps are well established, but preparing the atom pairs is an outstanding challenge. 


In this paper, we demonstrate the preparation of exactly one $^{87}$Rb atom and one $^{133}$Cs atom in the same optical tweezer as the essential first step towards the construction of a tweezer array of individually trapped $^{87}$Rb$^{133}$Cs molecules (hereafter RbCs). First, we detail the considerations that underlie the choice of tweezer wavelengths, and explain the importance of species-selectivity. Next, we show how we can prepare Rb and Cs atoms in separate, species-specific optical tweezers. Precise control of the position of each tweezer is essential for merging of the two tweezers for molecule formation. We demonstrate this control by overlapping the two tweezers and preparing one Rb and one Cs atom in the same tweezer with high fidelity. Finally, we use this capability to study hyperfine state dependent collisions of Rb and Cs in the combined tweezer, comparing the measured two-body loss rates with coupled-channel quantum scattering calculations. 




\section{Species-selective tweezers }
\label{sec1_wavelength_selection}

The preparation of heteronuclear atom pairs in a single optical tweezer, and the subsequent separation of the atoms back into their original tweezers for detection, are best achieved using species-selective optical tweezers. This selectivity can be engineered through a judicious choice of the tweezer wavelengths. Below we discuss the factors that influence our choice of wavelengths for the optical tweezers used to trap Rb and Cs.

The optical potential experienced by an atom in a far-off-resonant optical tweezer is described by \cite{Grimm2000}:
\begin{equation}
U_i(\lambda, r, z) = -U_{i,0}(\lambda)\times \frac{e^{    -2r^2 / (w_0^2 \tilde{z})}}{ \tilde{z}},
\end{equation}
where $U_{i,0}(\lambda)=\frac{1}{2 \epsilon_0 c} \alpha_i(\lambda) I_0$ is the species-dependent tweezer depth, $\lambda$ is the tweezer wavelength, $\alpha_i(\lambda)$ is the real part of the atomic polarisibility for species $i$, $I_0$ is the intensity at the focus of the tweezer, $w_0$ is the tweezer waist, $r$ is the radial coordinate, $z$ is the axial coordinate and $\tilde{z}=(1+(z/z_\mathrm{R})^2)$, with $z_\mathrm{R}=\pi w_0^2/\lambda$ the Rayleigh range. In the following discussion we use subscript notation to denote the wavelength of the tweezer used to trap each species ($\lambda_\mathrm{Rb}, \lambda_\mathrm{Cs}$) and the shorthand notation ($\tilde{U}_\mathrm{Rb}(\lambda), \tilde{U}_\mathrm{Cs}(\lambda)$) to denote the potential depths experienced by each species expressed in temperature units (i.e. $\tilde{U}_\mathrm{Rb}(\lambda)=U_{\mathrm{Rb},0}(\lambda)/k_\mathrm{B}$ and $\tilde{U}_\mathrm{Cs}(\lambda)=U_{\mathrm{Cs},0}(\lambda)/k_\mathrm{B}$).

\begin{figure}
    \centering
    \includegraphics[width=1\linewidth]{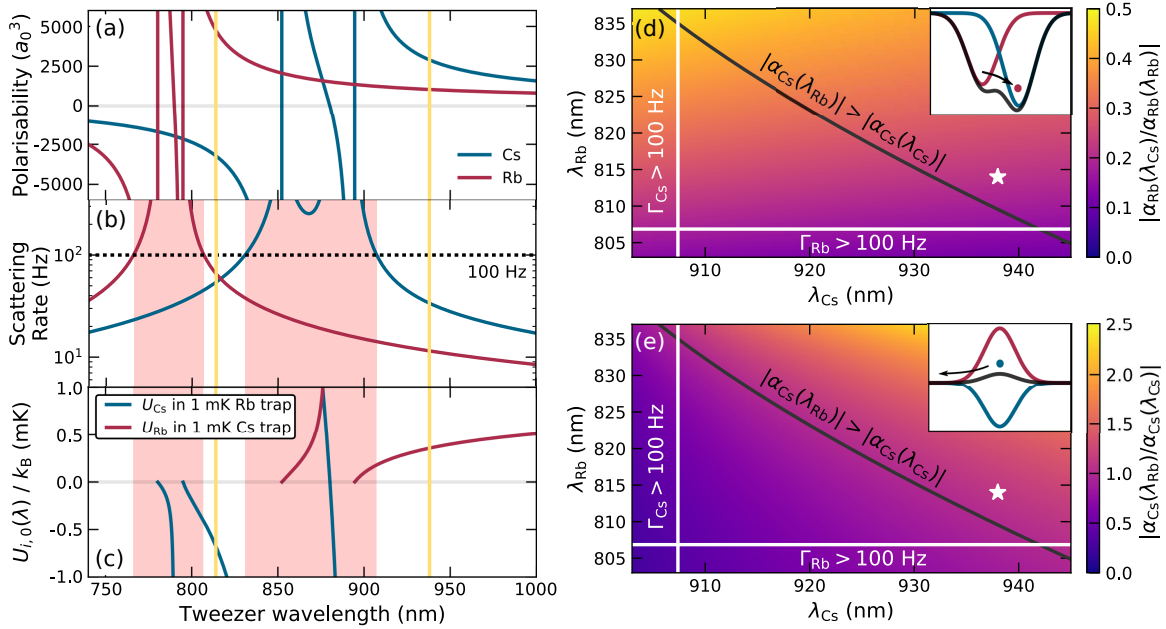}
    \caption{Factors influencing the wavelength selection for species-specific optical tweezers. (a) The wavelength-dependent polarisabilities for ground-state Rb and Cs atoms. (b) Photon scattering rates for Rb and Cs in a 1~mK deep tweezer. High scattering rates $> 100$~Hz (dotted line) preclude tweezer wavelengths in the red-shaded regions. (c) Potential depth experienced by one atomic species in a tweezer held at 1~mK depth for the other species. Only one suitable low-scattering wavelength band exists for each species. (d) Ratio of Rb polarisabilities as a function of the two tweezer wavelengths, $\lambda_\mathrm{Cs}$ and $\lambda_\mathrm{Rb}$. The inset illustrates the case where $\vert \alpha_\mathrm{Rb}(\lambda_\mathrm{Cs}) / \alpha_\mathrm{Rb}(\lambda_\mathrm{Rb}) \vert>1$, so that Rb spills into the deeper Cs tweezer during merging. (e) Ratio of Cs polarisabilities as a function of the two tweezer wavelengths, $\lambda_\mathrm{Cs}$ and $\lambda_\mathrm{Rb}$. The inset illustrates the case where $\vert \alpha_\mathrm{Cs}(\lambda_\mathrm{Rb}) / \alpha_\mathrm{Cs}(\lambda_\mathrm{Cs}) \vert>1$ so that the Cs atom is ejected by the antitrapping potential of the Rb tweezer during merging. In both (d) and (e), the black line indicates where $\alpha_\mathrm{Cs}(\lambda_\mathrm{Rb}) = \alpha_\mathrm{Cs}(\lambda_\mathrm{Cs})$ and the enclosed triangular regions correspond to the wavelength combinations with polarisability ratios that also satisfy the condition $\Gamma_{\mathrm{Rb,Cs}}<100$~Hz. The yellow lines in (a)--(c) and the white stars in (d) and (e) indicate the wavelengths used in the experiment. 
    }
    \label{fig1_polarisability}
\end{figure}

The atomic polarisibility is species- and wavelength-dependent, making it the critical parameter when engineering differential confinement between two species in the same optical tweezer. We calculate the polarisability for each species using methods presented in \cite{Safronova2006, Arora2007}. The polarisibilities for Cs in the $6^{2}$S$_{1/2}$ ground state (blue) and Rb in the $5^{2}$S$_{1/2}$  ground state (red) are shown in figure~\ref{fig1_polarisability}(a). The wavelength choice for each tweezer is first constrained by the sign of $\alpha_i(\lambda)$, which must be positive to realise an attractive potential. For each species, this is guaranteed for wavelengths red-detuned from the atomic D$_1$ transition which corresponds to tweezer wavelengths $\lambda_\mathrm{Rb}>795.0$~nm and $\lambda_\mathrm{Cs}>894.6$~nm.

Due to the micron-scale waist of an optical tweezer, high intensities at the focus ($\sim 1$~GW~cm$^{-2}$) can result in high photon scattering rates. The scattering rates for each species subjected to a tweezer with a depth of $\vert U_{i,0}(\lambda)\vert/k_{\mathrm{B}}=1$\,mK are shown in figure~\ref{fig1_polarisability}(b). We constrain the wavelength choice further by excluding regions where the scattering rate $\Gamma$ of a species exceeds 100~Hz (shaded red). We impose this limit so as to constrain the heating rate due to photon scattering to $\lesssim 20~\mu$K~s$^{-1}$ and to reduce the effect of off-resonant spontaneous Raman scattering, which can change the hyperfine state of the confined atom \cite{Cline1994}.  


By considering in more detail the case where the Rb and Cs tweezers are overlapped, we identify two further constraints on the choice of wavelength. First we require the Rb atom to experience a more confining potential in the Rb tweezer than in the Cs tweezer, i.e. that $\vert \tilde{U}_\mathrm{Rb}(\lambda_\mathrm{Rb}) \vert > \vert \tilde{U}_\mathrm{Rb}(\lambda_\mathrm{Cs}) \vert $. This is to avoid the Rb atom spilling into the Cs tweezer (illustrated in figure~\ref{fig1_polarisability}(d), inset) when the tweezers are merged. 
Secondly, we require that $\vert \tilde{U}_\mathrm{Cs}(\lambda_\mathrm{Cs}) \vert > \vert \tilde{U}_\mathrm{Cs}(\lambda_\mathrm{Rb}) \vert $ in the range of valid $\lambda_\mathrm{Rb}$ identified, so that the Cs atom is not expelled from its tweezer by repulsion from the Rb tweezer (as illustrated in figure~\ref{fig1_polarisability}(e), inset).

To understand how these two conditions can be satisfied, we now consider the potential experienced by each atom in the overlapped tweezers, using figures~\ref{fig1_polarisability}(c-e). Figure~\ref{fig1_polarisability}(c) illustrates the case where each atom experiences a confining potential of depth 1~mK from its own tweezer, i.e.\ $\tilde{U}_\mathrm{Cs}(\lambda_\mathrm{Cs})=1$~mK and $\tilde{U}_\mathrm{Rb}(\lambda_\mathrm{Rb})=1$~mK. The potential experienced by Rb due to the Cs tweezer $U_{\mathrm{Rb},0}(\lambda_\mathrm{Cs})$ 
is then shown in red, and the potential $U_{\mathrm{Cs},0}(\lambda_\mathrm{Rb})$ 
is shown in blue. For $\lambda_\mathrm{Cs}>907.3$~nm, 
$\vert \tilde{U}_\mathrm{Rb}(\lambda_\mathrm{Cs})\vert < 1$~mK, so that the Rb atom will not spill into the Cs tweezer.  Similarly, for $806.7~\mathrm{nm}<\lambda_\mathrm{Rb}<820.4~\mathrm{nm}$,  
$\vert \tilde{U}_\mathrm{Cs}(\lambda_\mathrm{Rb})\vert < 1$~mK, so that the Cs atom is not expelled from the Cs tweezer by the repulsion of the Rb tweezer. Figure~\ref{fig1_polarisability}(c) indicates that many choices of $\lambda_\mathrm{Cs}$ are valid, although it is desirable to use shorter wavelengths where smaller tweezer waists are achievable and lower laser powers are required.


The interplay between the two tweezers is further illustrated in figure~\ref{fig1_polarisability}(d) and figure~\ref{fig1_polarisability}(e), which show the polarisability ratios $\vert \alpha_\mathrm{Rb}(\lambda_\mathrm{Cs}) / \alpha_\mathrm{Rb}(\lambda_\mathrm{Rb}) \vert$ and $\vert \alpha_\mathrm{Cs}(\lambda_\mathrm{Rb}) / \alpha_\mathrm{Cs}(\lambda_\mathrm{Cs}) \vert$, respectively, as function of the two tweezer wavelengths. For clarity, we restrict the discussion to tweezers of equal powers and waists, so that the relative trap depths are simply determined by these polarisability ratios. The blue lines indicate the scattering rate limits and coincide with the shaded regions of figure~\ref{fig1_polarisability}(c). The black line indicates where $\alpha_\mathrm{Cs}(\lambda_\mathrm{Rb}) = \alpha_\mathrm{Cs}(\lambda_\mathrm{Cs})$. Above the line, $\vert \tilde{U}_\mathrm{Cs}(\lambda_\mathrm{Rb})\vert > \vert \tilde{U}_\mathrm{Cs}(\lambda_\mathrm{Cs})\vert$ and the Cs atom is ejected by the repulsive potential of the Rb tweezer during merging, as shown inset in figure~\ref{fig1_polarisability}(e). In contrast, for the wavelength ranges considered, the problem of spilling of the Rb atom into the Cs tweezer, shown inset in figure~\ref{fig1_polarisability}(d), is avoided as $\vert \tilde{U}_\mathrm{Rb}(\lambda_\mathrm{Rb})\vert > \vert \tilde{U}_\mathrm{Rb}(\lambda_\mathrm{Cs})\vert$. 
The enclosed triangular regions on each figure correspond to the wavelength combinations that fulfill all the conditions we have imposed.
The interplay between the tweezer wavelengths is clear in this 2D representation: for greater $\lambda_\mathrm{Cs}$, smaller $\lambda_\mathrm{Rb}$ is required to prevent expulsion of the Cs atom. We note that the enclosed region serves as useful guide to the choice of $\lambda_\mathrm{Rb}$ and $\lambda_\mathrm{Cs}$, but ultimately the tweezer powers used in the experiment are independently controllable, allowing wavelengths outside this region.

In the experiment we use tweezer wavelengths $\lambda_\mathrm{Rb}=814$~nm and $\lambda_\mathrm{Cs}=938$~nm. These are indicated by the yellow lines in figures~\ref{fig1_polarisability}(a)--(c) and the white stars in figures~\ref{fig1_polarisability}(d) and (e). We choose  $\lambda_\mathrm{Rb}=814$~nm to be close to the intersection of the Rb and Cs scattering rates shown in figure~\ref{fig1_polarisability}(b) and $\lambda_\mathrm{Cs}=938$~nm such that differential light shifts on the laser cooling transition were small~\cite{Hutzler2017}. In the 938~nm tweezer $\alpha_\mathrm{Cs}=2890~a_0^3$ and $\alpha_\mathrm{Rb}=1030~a_0^3$ (calculated using methods presented in \cite{Safronova2006}), so that the Cs atom experiences a potential a factor of 2.8 deeper than the Rb atom. In the 814~nm tweezer,  $\alpha_\mathrm{Cs}=-3220~a_0^3$ and $\alpha_\mathrm{Rb}=4760~a_0^3$. Although our chosen wavelengths lie outside the enclosed triangular region for equal intensity tweezers shown in figures~\ref{fig1_polarisability}(d) and (e), experimentally we use twice as much power in the 938~nm tweezer as in the 814~nm tweezer. This increases the confinement of the Cs atom sufficiently that it is not expelled by the Rb tweezer during merging.

\section{Experimental methods}
\label{sec:experimental_methods}
\begin{figure}
    \centering
    \includegraphics[width=0.6\linewidth]{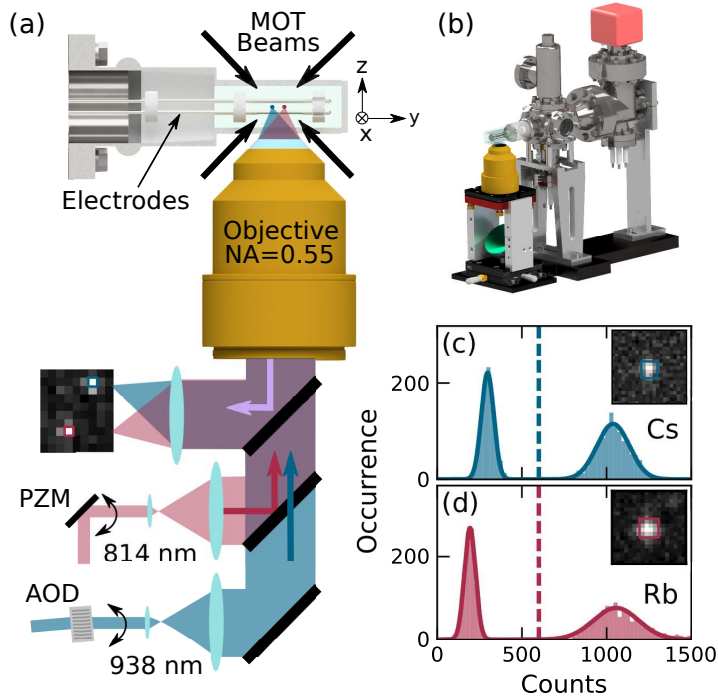}
    \caption{Trapping and imaging single Rb and Cs atoms. (a) Optical tweezers are focussed in to a science cell using a high-NA objective. 814~nm and 938~nm light are overlapped on a long-pass dichroic mirror and are then sent into the objective. Position control of the two tweezers is achieved using a piezo-controlled mirror (PZM) in the path of the 814~nm beam and an acousto-optic deflector (AOD) in the path of the 938~nm beam. 
    A second dichroic mirror picks off atomic fluorescence which is imaged on a camera (a typical image is shown). (b) 3D rendering of the vacuum apparatus and high-NA objective assembly. (c) A histogram of the fluorescence counts recorded when imaging Cs. The inset shows the atom image without pixel binning obtained from a single experimental run. The highlighted square shows the superpixel formed by binning 16 pixels. (d) A histogram of fluorescence counts from imaging Rb.}
    \label{fig2_imaging}
\end{figure}

\noindent
The starting point for the experiment is a dual 3D magneto-optical trap (MOT) of Rb and Cs prepared inside an UHV science cell, shown in figure~\ref{fig2_imaging}(a). The cell is attached to a single-chamber vacuum apparatus (figure~\ref{fig2_imaging}(b)) which contains the alkali-metal atom sources and mounts for the electrodes which, in the future, will be used to align the molecules in the laboratory frame  ~\cite{Gempel2016}. Each MOT is produced by three orthogonal beam pairs composed of overlapped cooling and repump light. Two beam pairs are indicated by the black arrows in figure~\ref{fig2_imaging}(a) and the third beam pair propagates orthogonal to the page (not shown). Each beam pair is formed by launching a beam from an optical fibre and retroreflecting it. The $1/e^2$ waists of the MOT beams are 1.5~mm and 1.6~mm for the Rb and Cs beams respectively. The Cs cooling light is red-detuned 8~MHz from the $f=4 \to f'=5$ free-space D$_2$ transition, and the Rb cooling light is red-detuned 13~MHz from the $f=2 \to f'=3$ free-space D$_2$ transition. Since we ultimately aim to trap only a few atoms in the optical tweezers, each MOT is typically loaded for just 150~ms to produce clouds of $1/e$ width $\sim 100$~$\mu$m containing less than $10^6$ atoms. 

The optical tweezers are formed using a high numerical aperture (NA = 0.55) objective lens (supplied by Special Optics) comprised of 7 optical elements which are designed to minimise the chromatic focal length shift at the imaging and tweezer wavelengths for both species. The effective focal length of the objective is 35.24~mm. The lens is located out of vacuum below the science cell, and is mounted to allow 3-axis translation and 2-axis angular control. 
The tweezer beams, at wavelengths 814~nm and 938~nm, are each derived from temperature-stabilised, free-running laser diodes which provide up to $\approx 10$~mW usable power at the atoms. A telescope in each beam path expands the beam diameter by a factor of 25 before the beams are overlapped on a long-pass dichroic mirror and focused into the science cell by the objective lens, where they overlap with the MOTs. The telescope spacing is adjusted to tune the tweezer beam divergence, in order to overlap the beam foci at the object plane. The positions of each tweezer in the science cell are independently controlled by a 2D piezo-controlled mirror (PZM) in the path of the 814~nm beam and an acousto-optic deflector (AOD) in the path of the 938~nm beam. 
This positional control is described in more detail in section~\ref{sec:merging}.

Atoms are loaded into the tweezers by loading the MOTs for 150~ms at a magnetic field gradient of 8.5~G~cm$^{-1}$, during which the tweezers are held at a 1~mK trap depth. The magnetic field gradient is then switched off and the frequency of the cooling beams is further detuned by $>50$~MHz to produce an optical molasses which cools the atoms further as they are loaded into the tweezer. Within the small-volume tweezer, collisional blockade ensures binary preparation of either zero or one atom in the trap \cite{Schlosser2002}. The loading probability is maximised when the MOT is well overlapped with the tweezer in 3D, which we optimise by adjusting shim fields applied along the $x$, $y$ and $z$ axes.


The high-NA objective lens is also used to image the trapped atoms. For imaging, the atoms are confined in a tweezer with a depth of 1~mK and illuminated by the MOT beams for 20~ms. The resulting atomic fluorescence is collected by the objective, split from the in-going tweezer beams using a custom dichroic mirror and then focussed onto an electron-multiplying charge-coupled device (EMCCD, Andor iXon Ultra 897) using an achromatic doublet lens. The overall detection efficiency of the imaging system is $\sim3$\,\%, largely determined by the $\sim8$\,\% collection efficiency associated with the solid angle of the objective lens. The magnification of the imaging system is 28.4, resulting in an effective pixel size at the EMCCD of $0.56~\mu$m. We reduce read-out noise by binning individual pixels into $4 \times 4$ superpixels. 


The stochastic loading of the optical tweezer, where either zero or one atom is loaded, results in either a low or a high number of counts detected during imaging. By repeating the same experimental sequence several hundred times at a typical rate of 2~Hz, a bimodal histogram of the atomic fluorescence is constructed \cite{Hu1994}. Histograms generated from 2000 runs of the experiment are shown for Cs and Rb in figures~\ref{fig2_imaging}(c) and (d), respectively. The insets show typical fluorescence images for each atomic species.  The loading probability into the tweezer is calculated from the ratio of images that exhibit counts falling above the threshold (dashed lines) to the number of images that exhibit counts below the threshold. Typically the atom loading probabilities for Rb and Cs are around $0.5$, consistent with other experiments that do not employ enhanced loading techniques \cite{Hutzler2017,Fung2015}. In a typical experiment, the probability of retaining an atom in the tweezer after fluorescence imaging is $>95$\,\%.


We employ a variety of established techniques to characterise the optical tweezers using the trapped atoms as a diagnostic. In a typical measurement, we first employ fluorescence imaging to verify atom occupancy and runs where no atom was loaded are discarded. The atom is then perturbed in some way to induce loss, for example by applying resonant light, and the tweezer is then probed for occupancy a second time to determine whether atom loss was induced. This is repeated over several hundred runs of the experiment to obtain an average survival probability and  error bars are calculated using a binomial confidence interval. By performing measurements in this manner, experimental observables can be mapped onto atom loss. For example, the energy distribution of the trapped atom is obtained by briefly turning off the tweezer between the first and second image and fitting the atom recapture probability to a Monte Carlo simulation \cite{Tuchendler2008}. For the experiments presented here, the fitted energy distribution corresponds to a temperature for each atom below $20~\mu$K, averaged over many iterations of the experiment. We measure the trap frequencies by parametric modulation of the tweezer intensity, which induces atom loss when the modulation frequency is equal to twice the trap frequency. The radial trap frequency $\nu_r$ is related to the tweezer waist by $w_0=\sqrt{4 U_{i,0}(\lambda)/(m (2 \pi \nu_r)^2)}$, from which we extract radial waists for each tweezer, $\{w_{x}^\mathrm{814}, w_{y}^\mathrm{814}\} = \{1.11(3), 0.92(3) \}~\mu$m and $\{w_{x}^\mathrm{938}, w_{y}^\mathrm{938}\} = \{1.29(4), 1.06(2) \}~\mu$m. The radial waists are along the $x$ and $y$ axes of the cell as defined in figure~\ref{fig2_imaging}(a) and the beam propagates along the $z$ direction. Asymmetric apodisation by beam-shaping optics before the objective breaks the cylindrical symmetry, yielding distinct radial beam waists along the $x$ and $y$ axes.


\section{Atom loading into species-selective optical tweezers}
We characterise  the species-selectivity of each optical tweezer by measuring the capture probability for Rb and Cs as a function of the tweezer power during the loading step. We typically use 200 runs of the experiment for each measurement. To maintain the same detection performance throughout the measurement, the tweezer power is always ramped to give a trap depth of 1~mK before imaging the atom.

\begin{figure}[t!]
    \centering
    \includegraphics[width=0.9\linewidth]{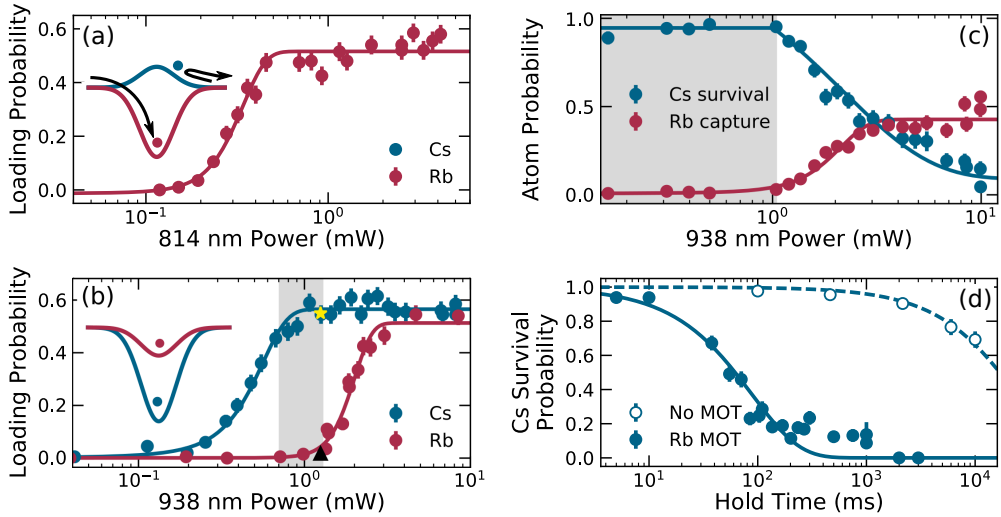}
    \caption{Atom loading into species-selective optical tweezers. (a) Rb loading probability as a function of 814~nm tweezer power. The 814~nm tweezer is repulsive for Cs (inset), so only Rb is loaded. (b) Loading probability of Rb and Cs as a function of 938~nm tweezer power. The tweezer is species-selective for the shaded band of powers. The yellow star and black triangle are 500-shot measurements of the loading into a 1.25~mW tweezer for Cs and Rb, respectively. The inset shows the relative potentials experienced by Rb (red) and Cs (blue). (c) Breakdown of species-selectivity in the 938~nm tweezer. A Rb MOT is overlapped for 500~ms with a tweezer containing a Cs atom. For tweezer powers $<1$~mW (shaded), the trap is species-selective. For higher powers, Rb is loaded into the trap causing loss of Cs by light-assisted collisions. (d) 
    Background-gas-limited lifetime of a Cs atom in an 8~mW non-selective 938~nm optical tweezer (open circles) compared to the survival time when overlapped with a Rb MOT (closed circles). Loading of Rb atoms reduces the Cs atom lifetime by a factor of $\sim270$. 
    }
    \label{fig3_species_selectivity}
\end{figure}

We first examine the species-selectivity of the 814~nm tweezer in figure~\ref{fig3_species_selectivity}(a) by measuring the atom loading probability as a function of the tweezer power. As expected, since the 814~nm tweezer potential is repulsive for Cs (see inset), no Cs atoms load at any power and the tweezer is perfectly species-selective. The Rb loading probability saturates at 0.52(1) for powers $> 0.5$~mW, corresponding to a trap depth of 0.33(2)~mK. 

The behaviour of the 938~nm tweezer is very different, as it is attractive for both Rb and Cs. The loading probabilities for Rb (red) and Cs (blue) into the 938~nm tweezer as a function of tweezer power, $P_{938}$, are shown in figure~\ref{fig3_species_selectivity}(b). The Cs loading probability saturates at a lower power than the Rb loading probability because Cs experiences a deeper confining potential in the 938~nm tweezer. The Cs loading probability saturates at 0.57(1) for tweezer powers greater than 1~mW, corresponding to a trap depth of 0.30(2)~mK. The Rb loading probability into the 938~nm tweezer saturates to 0.51(3) for powers greater than 3~mW, corresponding to a trap depth of 0.32(2)~mK, in agreement with the Cs measurement. We note that the loading probabilities for Rb and Cs into both tweezers saturate at similar trap depths of around 0.32~mK due to the similar temperatures and densities of the MOTs. The 938~nm tweezer is species-selective for powers in the range $0.7$~mW~$<P_{938}<1.3$~mW (shaded grey), where the Cs loading probability is close to saturation and the Rb loading probability is close tpecifically, for a tweezer power of 1.25~mW, we measure loading probabilities of 0.018(6) for Rb (black triangle) and 0.55(2) for Cs (yellow star) over 500 repetitions of the experiment. At this power, the tweezer can therefore be used to load a Cs atom selectively. 

For powers greater than 1.3~mW in the 938~nm tweezer beam, the loading probabilities of Rb and Cs both saturate, so the tweezer is no longer species-selective. We examine this breakdown in species-selectivity in  figure~\ref{fig3_species_selectivity}(c). Here, a Cs atom is first loaded into the 938~nm tweezer and the trap is checked for occupancy, after which the tweezer power is ramped to a variable value. A Rb MOT is then loaded, overlapping with the tweezer for a time of 500~ms before an optical molasses stage. The occupancy of the 938~nm tweezer is then probed to measure the survival of Cs atoms (blue points) and to check for the capture of Rb atoms (red points). 
For tweezer powers less than 1~mW (shaded grey), the tweezer is species-selective so that no Rb atoms are loaded and the Cs survival probability is high. At greater tweezer powers, we observe the capture of Rb atoms correlated with a reduction in the Cs survival probability.
We believe that the exchange of Rb for Cs in the tweezer happens in two steps. Firstly, if a Rb atom is loaded into the tweezer containing a Cs atom, interspecies light-assisted collisions in the presence of the near-resonant MOT light~\cite{Telles2001,Harris2008} can lead to loss of both atoms. Subsequently, a Rb atom can be loaded into the empty tweezer. We have independently determined that the $1/e$ loading time for single Rb atoms into a 1~mK (9.4~mW) tweezer is 60(20)~ms. Since we overlap the Rb MOT with the tweezer for 500~ms in this measurement, there is sufficient time for another Rb atom to load into the tweezer, reproducing the behaviour observed in figure~\ref{fig3_species_selectivity}(c). 

The loading of Rb atoms drastically reduces the lifetime of the Cs atom in the 938~nm tweezer, as shown in figure~\ref{fig3_species_selectivity}(d). The open circles show the survival probability of a single Cs atom in the absence of the Rb MOT. We extract a $1/e$ lifetime of 24(3)~s, limited by collisions with background gases in the science cell. The solid circles show the impact on the single-atom lifetime when the Rb MOT is loaded and maintained during the variable hold time. The tweezer is held at power of 8~mW, which is strongly non-selective. The Cs $1/e$ survival time is reduced to 90(10)~ms by collisions with Rb atoms, a factor of $\sim270$ decrease in survival time.

Following this discussion, two possible loading schemes become evident: the Rb and Cs MOTs can be prepared simultaneously, to load the 814~nm tweezer and the 938~nm tweezer at the same time, or the MOTs can be prepared in succession, to load the tweezers sequentially. The appeal of the first case is that, by eliminating the 150~ms time required to load a second MOT separately, the data collection rate can be increased by $\sim50\,\%$. However, even for a species-selective 938~nm tweezer there is $2\,\%$ chance of loading a Rb atom. Furthermore, due to technical limitations it is challenging to overlap both MOTs with the optical tweezers simultaneously using the same shim magnetic fields. We therefore opt for the second case, first loading a Rb atom into the 814~nm tweezer, followed by a Cs atom into the 938~nm tweezer. Although this scheme is slower, the repulsive nature of the 814~nm tweezer towards Cs precludes any cross-loading of an atom into the wrong tweezer. Operating both tweezers in the regime where the loading probabilities are saturated, we performed the sequential loading routine over 6000 runs of the experiment. Using this sequence, we measured loading of single Rb atoms into the 814~nm tweezer in 50.8(6)\,\% of runs, single Cs atoms into the 938~nm tweezer in 54.7(6)\,\% of runs, and the preparation of exactly one Rb and one Cs atom in 28.4(6)\,\% of runs. 

\section{Controlling and merging species-selective tweezers}
\label{sec:merging}
\noindent
We now explore the merging of two tweezers, each containing a single atom, into a single tweezer containing both atoms. Following the loading, our goal is to move the 938~nm tweezer (containing Cs) to overlap the stationary 814~nm tweezer (containing Rb) such that the power of the 814~nm tweezer can be reduced to zero, transferring the Rb atom into the 938~nm tweezer. Efficient transfer with minimal heating requires good overlap between the two tweezers along the two radial directions, $x$ and $y$, and the axial direction, $z$, as defined in figure~\ref{fig2_imaging}(a).



\begin{figure}
    \centering
    \includegraphics[width=1\linewidth]{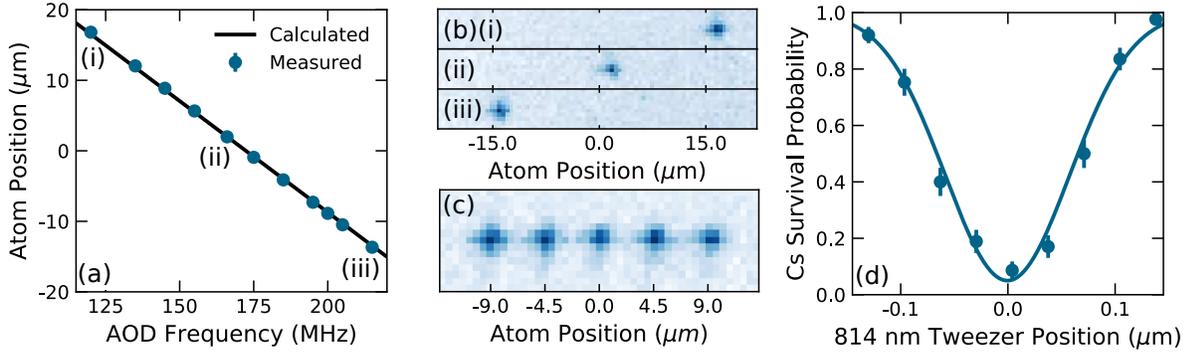}
    \caption{Control of the optical tweezer positions. (a) Position of the 938~nm tweezer as a function of the RF frequency applied to the AOD. The black line shows the displacement expected using equation (\ref{eq_control}). The solid points show the measured atom displacements.  (b) Images of Cs atoms in displaced optical tweezers. The images (i)-(iii) correspond to the labelled points in (a). (c) Five optical tweezers are generated by simultaneously driving the AOD at five frequencies. Cs atoms load into each optical tweezer. The image is an average of 200 experimental runs. (d) Overlap of the 814~nm tweezer with the 938~nm tweezer using the piezo-controlled mirror. Loss of a trapped Cs atom is induced by the repulsive 814~nm potential with the power purposely chosen to induce loss when the tweezers are overlapped.}
    \label{fig4_control}
\end{figure}


The AOD in the 938~nm beam path is used to control of the position of the tweezer along $x$. The RF signal applied to the AOD is generated by an arbitrary waveform generator with a sample rate of $6.25\times 10^8$~samples/s (Spectrum Instrumentation M4i.6622-x8). The diffraction angle $\theta$ of the first-order transmitted beam is given by $\sin\theta = \lambda f_\mathrm{RF}/v_\mathrm{s}$, where $\lambda=938$~nm, $v_\mathrm{s}$ is the speed of sound in the acoustic crystal and $f_\mathrm{RF}$ is the frequency of the applied RF tone. By dynamically sweeping the frequency of the RF tone, the diffraction angle can be adjusted during an experimental cycle, allowing control of the tweezer position. For a frequency sweep $\Delta f_\mathrm{RF}$, the corresponding tweezer displacement at the plane of the atoms is given by
\begin{equation}
    \Delta x = \frac{\lambda f_\mathrm{obj}}{v_\mathrm{s} M_\mathrm{tel}} \Delta f_\mathrm{RF},
\label{eq_control}
\end{equation}
where $f_\mathrm{obj}$ is the effective focal length of the objective and $M_\mathrm{tel}=25$ is the magnification of the tweezer expansion telescope in figure~\ref{fig2_imaging}(a). 

We have chosen a longitudinal-mode AOD with a high speed of sound $v_s=4200$~m~s$^{-1}$ (IntraAction ATD-1803DA2.850). This choice was dictated by the long-term goal of preparing RbCs molecules in a tweezer array where the tweezer spacing is $\lesssim 1~\mu$m in order to achieve dipole-dipole interaction energies of $\sim h \times 1$~kHz between neighbouring molecules~\cite{Blackmore2019}. This close spacing requires the AOD to be driven with several tones close in frequency, resulting in a frequency beat note and modulation of the trapping potential. If this modulation frequency is comparable to the trap frequencies, parametric heating of the trapped atoms or molecules will occur \cite{Endres2016}. The high speed of sound in our chosen AOD means that a tweezer separation of $1~\mu$m at 938~nm still requires a frequency spacing of $\sim 3$~MHz between the RF tones. This is sufficiently large with respect to the trap frequencies ($\lesssim 100$~kHz) that parametric heating will be negligible.

Control of the tweezer position with AOD frequency is demonstrated in figure~\ref{fig4_control}(a). The data points show the displacement of the tweezer determined by fitting the images of single atoms (figure~\ref{eq_control}(b)) with a Gaussian function to extract the centre. The solid line shows predicted displacement of the tweezer according to equation~(\ref{eq_control}). 
The fitted dependence of the tweezer displacement on the RF frequency is 0.322(1)~$\mu$m~MHz$^{-1}$. The bandwidth of the AOD is 90~MHz, so that the tweezer position can be scanned up to $29~\mu$m in the $x$ direction to be overlapped with the 814~nm tweezer. 

The scalability of our approach is demonstrated in figure~\ref{fig4_control}(c), which shows an image of an array of 5 Cs atoms, averaged over 200 runs of the experiment. The AOD is driven simultaneously by 5 RF tones spaced by 14~MHz in frequency, generating an array of tweezers with spacing 4.5~$\mu$m. We equalise the trap depths to within $2~\%$ by modifying the RF amplitude of each tone iteratively, conditional on the measured trap frequency of each tweezer. In the future, we plan to apply 
rearrangement techniques \cite{Lee2016, Barredo2016, Endres2016} to such an array to increase the fraction of experimental runs where both a Rb and a Cs atom are loaded.

The tweezer overlap in the $y$ direction is achieved using the piezo-controlled mirror (PZM) in the 814~nm beam path. The PZM angle is ratcheted in discrete steps controlled by an open-loop driver. Each voltage step increments the mirror angle by 0.14 arcseconds. In the atom plane, the measured displacement of the tweezer is 1.34(7)~$\mu$m per 1000 steps. To measure the overlap of the tweezers in the $y$ direction, we exploit the repulsion of Cs by the 814~nm tweezer. After a Cs atom is loaded into the 938~nm tweezer, the AOD is used to sweep the position of the 938~nm tweezer through 10~$\mu$m and back to the starting position in 40~ms, translating it through the $x$-coordinate of the 814~nm tweezer. The PZM angle is varied so that the $y$ position of the 814~nm tweezer is stepped through the scanning 938~nm tweezer. By setting the tweezer powers such that $\vert \tilde{U}_\mathrm{Cs}(\lambda_\mathrm{814})\vert>\vert \tilde{U}_\mathrm{Cs}(\lambda_\mathrm{938})\vert$, we ensure that the Cs atom in the 938~nm tweezer is expelled when the tweezers overlap, as shown in figure~\ref{fig4_control}(d). The PZM is set to give the optimal $y$ overlap, which is known from the fit to the signal in figure~\ref{fig4_control}(d) to within a 1$\sigma$ uncertainty of 20~nm.


\begin{figure}
    \centering
    \includegraphics[width=1\linewidth]{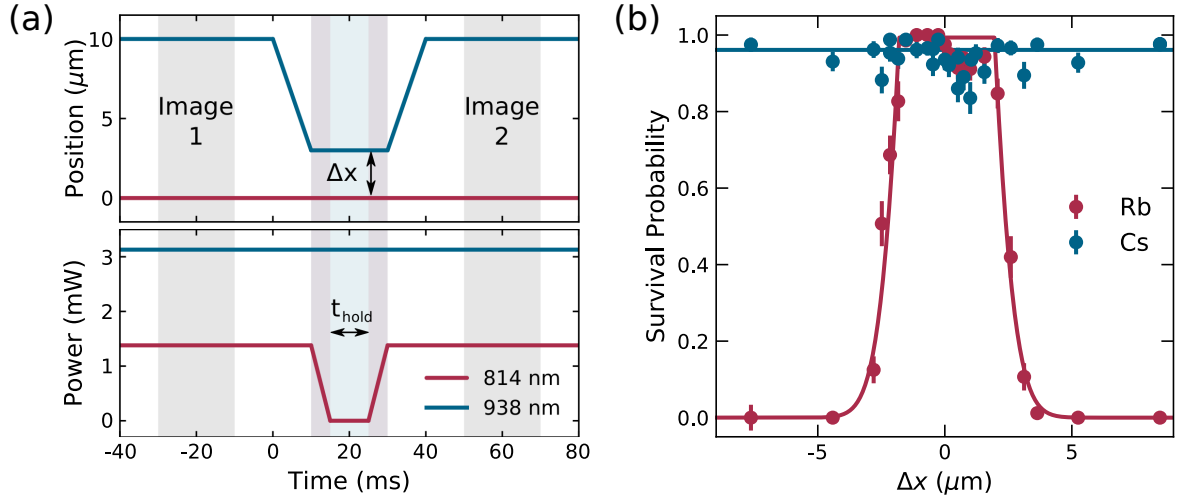}
    \caption{Merging Cs and Rb into a single optical tweezer. (a) Timing diagram for the tweezer merging sequence. After both tweezers are checked for atom occupancy (Image 1), the AOD frequency is ramped to overlap the 938~nm optical tweezer with the 814~nm tweezer (red-shaded region). The 814~nm tweezer is ramped off so that both atoms are confined in the 938~nm tweezer (blue-shaded region). After a hold time, the sequence is reversed, and the atom occupancy is again probed (Image 2). (b) Survival probability of Cs in the 938~nm tweezer and Rb in the 814~nm tweezer after the merging sequence in (a) for a variable $\Delta x$ and $t_\mathrm{hold}=10$~ms. For a range $-2~\mu$m~$< \Delta x < 2~\mu$m there is good transfer of the Rb atom between the tweezers.}
    \label{fig:5_merging}
\end{figure}


Overlap in the $z$ direction is performed by repeating the measurement in figure~\ref{fig4_control}(d) for several axial displacements of the focus of the 814~nm tweezer. The $z$ position of the 814~nm focus can be tuned by adjusting the divergence of the beam as it passes through the $M=25$ telescope in figure~\ref{fig2_imaging}(a). When the axial overlap of the two tweezers is optimal, the depth of the loss feature in figure~\ref{fig4_control}(d) is maximised since the increase in 814~nm intensity causes a stronger repulsion. The Rayleigh range of the 814~nm tweezer is $3~\mu$m, and the axial position of the 814~nm tweezer is set to within 0.5~$\mu$m of the 938~nm tweezer. We have observed an hour-scale drift in the radial overlap of less than 150~nm, and a 200~nm drift over one month.

With the tweezers aligned in the $y$ and $z$ axes, we perform merging of Rb and Cs into the same tweezer along the $x$ direction using the experimental sequence shown in figure~\ref{fig:5_merging}(a). The tweezers are loaded and imaged to check for atom occupancy (shaded region labelled Image~1). The AOD frequency is set to 140~MHz and the 938~nm tweezer power is held at 3.13~mW. The 814~nm tweezer is loaded at a power of 1.38~mW. For these powers, the potential depths experienced by each species are: $\tilde{U}_\mathrm{Cs}(\lambda_{938})=0.95$~mK, $\tilde{U}_\mathrm{Cs}(\lambda_{814})=-0.62$~mK, $\tilde{U}_\mathrm{Rb}(\lambda_{814})=0.92$~mK and $\tilde{U}_\mathrm{Rb}(\lambda_{938})=0.34$~mK, yielding combined trap depths of $\tilde{U}_\mathrm{Cs, merged}=0.33$~mK and $\tilde{U}_\mathrm{Rb, merged}=1.26$~mK. Crucially, for the 814~nm power used, the anti-trapping potential applied to Cs is insufficient to eject it from the 938~nm tweezer when the tweezers are overlapped. The AOD frequency is then swept to a target frequency in 10~ms, translating the $x$ position of the 938~nm tweezer to a distance $\Delta x$ from the 814~nm tweezer. The power of the 814~nm tweezer is then adiabatically ramped to zero in $5$~ms. If the overlap between the tweezers is good, the Rb atom is transferred into the 938~nm tweezer; if there is insufficient overlap, the Rb atom is lost and not recaptured into the 814~nm tweezer. After a time $t_\mathrm{hold}=10$~ms, the sequence is reversed and the occupancy of each tweezer is probed again (shaded region labelled Image~2). Due to the species-selectivity of each tweezer, when the tweezers are separated, each atom returns to its original tweezer. In figure~\ref{fig:5_merging}(b) we present the Cs and Rb single atom survival probabilities as a function of $\Delta x$. When the tweezers are overlapped to within $2~\mu$m along $x$, the Rb atom is retained with $>0.90$ probability. At the optimal overlap, which occurs at an AOD frequency of 163.3~MHz, the Cs survival probability is $0.99^{(+0.01)}_{(-0.02)}$ and the Rb survival probability is $1.00^{(+0.00)}_{(-0.01)}$, yielding a combined merging and separation survival probability of $0.99^{(+0.01)}_{(-0.02)}$.

We determine the two-atom pair density for Rb and Cs in the 938~nm tweezer following merging using expressions given in \cite{Anderlini2005}. We independently measure temperatures of $T_\mathrm{Cs}=10(3)~\mu$K and $T_\mathrm{Rb}=15(5)~\mu$K following the optimal merging sequence. These are within error of the initial values indicating little heating from the merging process. We also measure the trap frequencies for each atom in the merged tweezer by parametric heating. The Cs trap frequencies are \{$\nu_{x}, \nu_{y}, \nu_{z}\} = \{54(5), 80(5), 12(1)\}$~kHz and the Rb trap frequencies are \{$\nu_{x}, \nu_{y}, \nu_{z}\} =   \{40(4), 59(3), 9(2)\}$~kHz. 
For these parameters, we calculate a pair density of $n_\mathrm{Rb,Cs}=5(2)\times 10^{12}$~cm$^{-3}$.





\section{State-dependent Rb + Cs collisions}

The case of two atoms in a single optical tweezer is a near-pristine environment to study two-body collisions. One-body losses can be neglected due to the long vacuum lifetime, and the three-body losses often seen in bulk-gas experiments are entirely suppressed by the exact atom number control in the tweezer. Here we investigate two-body loss resulting from hyperfine-changing collisions between Cs and Rb in the 938~nm tweezer. The hyperfine energy splittings in the ground state of $^{87}$Rb and Cs are $h\times6.8$\,GHz and $h\times9.2$\,GHz, respectively. If at least one atom is in the upper hyperfine state, then hyperfine-changing collisions can convert this energy (equivalent to $>100$~mK, far in excess of the typical $\sim 1$~mK trap depths) into kinetic energy shared between the two atoms \cite{Ueberholz2002}, leading to loss of both atoms from the tweezer.


\begin{figure}
    \centering
    \includegraphics[width=0.87\linewidth]{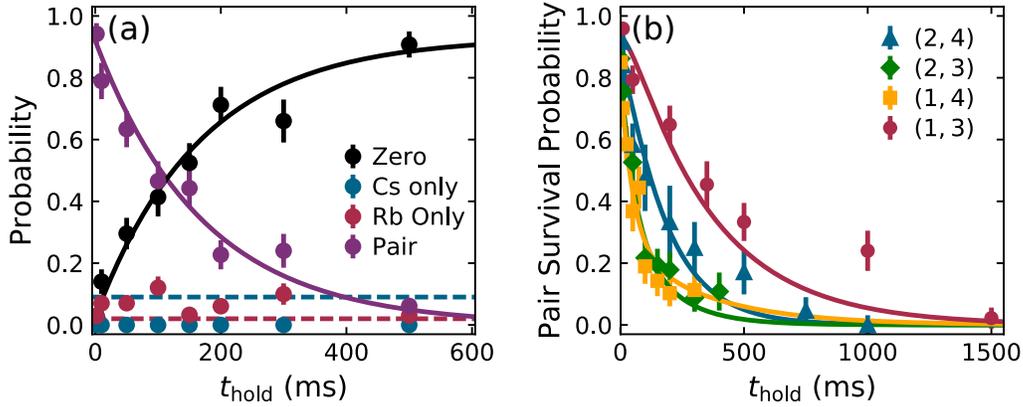}
    \caption{Rb and Cs collisions in an optical tweezer. (a) Rb and Cs atoms are merged into a single tweezer, without optical pumping, for a variable time $t_\mathrm{hold}$. 
    The pair survival probability (purple) is post-selected for events when a Rb and Cs atom are both loaded. The probability of measuring no atoms after $t_\mathrm{hold}$ is shown in black. The probability of observing just Rb or Cs is shown in red and blue respectively.  The dashed lines indicate $1-P_1$, where $P_1$ is the single-atom survival probability, post-selected on runs where only a Rb (red) or Cs (blue) atom was loaded. (b) Pair survival probability of optically pumped Rb and Cs atoms, post-selected for pair loading events. Before the tweezers are merged, the atoms are first optically pumped to the hyperfine combinations $(f_\mathrm{Rb},f_\mathrm{Cs}) = (2,4)$ (blue triangles), $(2,3)$ (green diamonds), $(1,4)$ (yellow squares) and $(1,3)$ (red circles). The solid lines are fitted to a coupled rate model as described in the text. }
    \label{fig6_collisions}
\end{figure}

We probe the rate of hyperfine-changing collisions by performing the merge sequence described in section~\ref{sec:merging}, varying the time $t_\mathrm{hold}$ for which the atoms are held together in the 938~nm tweezer before separation. There are four possible scenarios when the experiment is initialised: either zero atoms are loaded, one Rb is loaded, one Cs is loaded, or one Rb \textit{and} one Cs atom are loaded, occurring with probabilities $P= 0.229(5),~0.224(5),~0.263(6),~0.284(6)$ respectively, measured over 6000 runs of the experiment. We post-select experimental runs where one Rb and one Cs atom were present in the first fluorescence image. Figure~\ref{fig6_collisions}(a) shows the pair loss for a Cs and Rb pair prepared in a mixture of hyperfine and Zeeman states. Stray external magnetic fields are cancelled to $<0.1$~G and no bias field is applied, in order to measure a collision rate averaged over $m_f$ states. Loss due to $m_f$-changing collisions can be neglected since the energy splitting of Zeeman substates is much smaller than the trap depth. The pair survival probability decreases exponentially, with a commensurate increase in the probability of observing no atoms after the tweezers are separated. The non-zero probabilities of losing a single Cs or Rb atom are due to artefacts during fluorescence imaging. In this measurement, for Cs we find there is $9(4)\%$  loss during the imaging, leading to the background rate of single Rb atoms shown in figure 6(a). The dashed lines indicate $1-P_1$, where $P_1$ is the single-atom survival probability post-selected on runs where only a Cs (blue) or Rb (red) atom was present.


We now examine the loss rates for optically pumped atom pairs. Hyperfine-state optical pumping is achieved using the MOT beams to apply a $7$~ms pulse of either cooling or repump light to each atom before the tweezers are merged. This allows the Cs atom to be prepared in either $6^{2}$S$_{1/2}$ $f_\mathrm{Cs}=3$ or $f_\mathrm{Cs}=4$, and the Rb atom to be prepared in either $5^{2}$S$_{1/2}$ $f_\mathrm{Rb}=1 $ or $f_\mathrm{Rb}=2$. We characterise the optical pumping fidelity by using a state-sensitive detection scheme: after state preparation, the tweezer depth is lowered to $300~\mu$K and a pulse of light resonant with the cooling transition is applied for $100~\mu$s, pushing out any population in the upper $f$ state. For optical pumping to both upper and lower hyperfine states, we measure fidelities greater than $99 \%$ at short experimental hold times for both Rb and Cs. 

Optical pumping of both species gives rise to one of four possible hyperfine combinations for the atom pair:  $\{(f_\mathrm{Rb},~f_\mathrm{Cs})\} = \{(2, 4); \ (2, 3);  \ (1, 4); \ (1, 3) \}$. We find that, for the hold times used in the experiment, spontaneous Raman scattering~\cite{Cline1994, Ozeri2005} due to the intense 938~nm tweezer light leads to redistribution between the hyperfine states. We therefore analyse the loss of atom pairs using a coupled rate model, 
\begin{equation}
\frac{d}{dt}
\begin{pmatrix}
      P_{24} \\
      P_{23}\\
      P_{14}  \\
      P_{13} \\
    \end{pmatrix}  =
    \begin{pmatrix}
      -\Gamma_{24}-r_\mathrm{Cs}-r_\mathrm{Rb} & r_\mathrm{Cs}& r_\mathrm{Rb}&0 \\
      r_\mathrm{Cs} &   -\Gamma_{23}-r_\mathrm{Cs}-r_\mathrm{Rb}& 0&r_\mathrm{Rb} \\
      r_\mathrm{Rb}  & 0& -\Gamma_{14}-r_\mathrm{Cs}-r_\mathrm{Rb} &r_\mathrm{Cs}\\
      0  & r_\mathrm{Rb}& r_\mathrm{Cs}& -r_\mathrm{Cs}-r_\mathrm{Rb}\\
    \end{pmatrix}
\begin{pmatrix}
      P_{24} \\
      P_{23}\\
      P_{14}  \\
      P_{13} \\
    \end{pmatrix}  ,
\label{eq:rate_equation}
\end{equation}
where for example $P_{24}(t)$ is the survival probability of the $(2,4)$ atom pair after a time $t$, and $\Gamma_{24}$ is the pair loss rate. 
The loss rate $\Gamma_{13}$ does not appear because there are no hyperfine-changing collisions when both atoms are in the lower ground-state manifold.  

We include the fraction of spontaneous Raman scattering events which cause a change  of $f_\mathrm{Rb}$ or $f_\mathrm{Cs}$ in our model by the addition of the rates $r_\mathrm{Rb}$ and $r_\mathrm{Cs}$. We have independently measured these rates for Rb and Cs in the 938~nm tweezer. We find $r_\mathrm{Cs} =3.9(7)$~Hz and use this to constrain the rate model. The spontaneous Raman scattering rate for Rb is very low due to the greater detuning of the tweezer wavelength from the atomic transitions. Our measurement of $r_\mathrm{Rb}$ is limited by the single-atom lifetime, indicating an upper limit of $r_\mathrm{Rb}<0.02$~Hz. The true scattering rate is expected to be negligible, so we set $r_\mathrm{Rb}=0$ in equation (\ref{eq:rate_equation}). We note that calculations using the method of \cite{Ozeri2005} indicate that $r_\mathrm{Cs}$ could be reduced by a factor of $100$ if a (less species-selective) $1064$~nm tweezer were used instead.

\begin{table}
\caption{\label{tab:k2s}Two body loss rates for Rb and Cs collisions in a merged optical tweezer. The experimental values are extracted from a rate equation fit to measured data, from which $k_2$ values are calculated using the effective pair density. The theory values are obtained from coupled-channel calculations as described in the text. 
}
\begin{indented}
\item[]\begin{tabular}{@{}crcc}
\br
$(f_\mathrm{Rb},~f_\mathrm{Cs})$ & $\Gamma/2\pi$ (s$^{-1}$) & $k_2$(expt)~(cm$^3$~s$^{-1})$ & $k_2$(theory)~(cm$^3$~s$^{-1})$\\
\mr
$(2,~4)$ &  4(1) &  $9(4) \times 10^{-13}$ &  $1.419 \times 10^{-12}$ \\
$(2,~3)$ & 12(1) &  $3(1) \times 10^{-12}$ &  $1.118 \times 10^{-12}$ \\
$(1,~4)$& 14(1) &  $3(1) \times 10^{-12}$  &  $2.347 \times 10^{-12}$ \\

\br
\end{tabular}
\end{indented}
\end{table}


The measured pair survival probability after a variable hold time in the 938~nm tweezer following merging is shown in figure~\ref{fig6_collisions}(b) for each hyperfine-pair combination. We fit $\Gamma_{24}, \ \Gamma_{23}$ and $\Gamma_{14}$ simultaneously to equation (\ref{eq:rate_equation}), obtaining the solid lines in figure~\ref{fig6_collisions}(b). 
The extracted loss rates are summarised in table~\ref{tab:k2s} for each hyperfine combination. The corresponding two-body loss rate constants, given by $k_2= \Gamma/ n_\mathrm{Rb,Cs}$, are also given. Our measured $k_2$ values are of a similar order of magnitude to those estimated from other (non-tweezer based) experiments on Rb $+$ Cs collisions \cite{Gibbs1967, Anderlini2005_2}. As noted above, collisional loss of pairs prepared in $(1,3)$ is energetically forbidden. However population can leak from this pair state by spontaneous Raman scattering to other pair combinations where hyperfine-changing collisions are allowed, resulting in the slower observed pair loss rate from $(1, 3)$. We independently fit the $(1,3)$ combination, yielding a $1/e$ time of $0.5(1)$~s. Since magnetoassociation to form RbCs molecules from atoms in $(f_\mathrm{Rb}=1,m_{f,\mathrm{Rb}}=1)$ and $(f_\mathrm{Cs}=3,m_{f,\mathrm{Cs}}=3)$ requires less than 10~ms, spontaneous Raman scattering from the tweezer is not expected to present an obstacle to molecule formation.


\section{Coupled-channel scattering calculations}

In the absence of external fields, the individual atoms in a tweezer do not have conserved projection quantum numbers $m_f$. Under these circumstances, $f_\textrm{Rb}$ couples to $f_\textrm{Cs}$ to form a resultant $F$. If spin relaxation due to magnetic dipolar and second-order spin-orbit coupling is neglected, $F$ is conserved in collisions at zero field. Degeneracy-averaged inelastic rate coefficients can be calculated as described by Xu et al.\ \cite{Xu2015}.

In an optical trap or tweezer, motional states for different values of $F$ would have slightly different interaction shifts because the scattering length depends on $F$. However, such differences are likely to be on the order of a few kHz, and a magnetic field of even a few mG is sufficient to decouple $f_\textrm{Rb}$ and $f_\textrm{Cs}$. Under these circumstance the initial and final states for collisions have well-defined $(f_\textrm{Rb},m_{f,\textrm{Rb}})$ and $(f_\textrm{Cs},m_{f,\textrm{Cs}})$. Collisions can change these quantum numbers, while conserving $M_F=m_{f,\textrm{Rb}}+m_{f,\textrm{Cs}}$ (except for spin relaxation as above). Collisions that change $m_{f,\textrm{Rb}}$ and $m_{f,\textrm{Cs}}$ without changing $f_\textrm{Rb}$ or $f_\textrm{Cs}$ do not release sufficient kinetic energy to eject atoms from the tweezer, unless the magnetic field is more than a few G, but they do redistribute population between different $(m_{f,\textrm{Rb}},m_{f,\textrm{Cs}})$ pairs with the same $M_F$.

\begin{table}
\centering
\caption{Calculated degeneracy-averaged inelastic rate coefficients for
$(f_\textrm{Rb},f_\textrm{Cs}) = (2,3) \rightarrow (1,3)$ as a function
of $M_F$ at $E/k_\textrm{B}=20\ \mu$K and limitingly low magnetic field. }
\begin{indented}

\item[]\begin{tabular}{@{}crrrrrc}
\br 
$M_F$ & \multicolumn{5}{c}{Contributing $m_{f,\textrm{Rb}},m_{f,\textrm{Cs}}$} & $k^{M_F}_{2,3}~(10^{-12}$ cm$^3$\,s$^{-1} )$\\
\mr 
0   & -2,2 & -1,1 & 0,0 & 1,-1 & 2,-2 & 1.039 \\
1   & -2,3 & -1,2 & 0,1 & 1,0  & 2,-1 & 1.039 \\  
2   &      & -1,3 & 0,2 & 1,1  & 2,0  & 1.298 \\
3   &      &      & 0,3 & 1,2  & 2,1  & 1.428 \\
4   &      &      &     & 1,3  & 2,2  & 1.153 \\
5   &      &      &     &      & 2,3  & 0 \\
all &      &      &     &      &      & 1.118 \\
\br
\end{tabular}
\label{table:rates-MF}
\end{indented}
\end{table}

We carry out coupled-channel quantum scattering calculations of collisions between unconfined Rb and Cs atoms using the MOLSCAT package \cite{molscat:2019, mbf-github:2020} with the methods and interaction potentials described in ref.\ \cite{Takekoshi2012}. These produce inelastic cross sections between individual pair states $(f_\textrm{Rb},m_{f,\textrm{Rb}}) + (f_\textrm{Cs},m_{f,\textrm{Cs}})$ as a function of relative collision energy $E$ and magnetic field $B$. For each value of $M_F$, we multiply the cross sections by the relative velocity and average over initial $m_{f,\textrm{Rb}}$ and $m_{f,\textrm{Cs}}$ to produce rate coefficients $k^{M_F}_{f_\textrm{Rb},f_\textrm{Cs} \rightarrow f_\textrm{Rb}',f_\textrm{Cs}'}$. These are in turn summed over final states to produce total loss rate coefficients $k^{M_F}_{f_\textrm{Rb},f_\textrm{Cs}}$. These are still labelled by $M_F$, which is a conserved quantity for each atom pair when spin relaxation is neglected.

Atomic Zeeman splittings are significant compared to the collision energy even at magnetic fields below 0.1~G. Our computational method can take account of such effects, but they obscure the physical picture. To avoid them, we present here results at limitingly low field. For the purpose of illustration, we consider relaxation from $(f_\textrm{Rb},f_\textrm{Cs}) = (2,3)$, where the only loss channels are those that produce $(f_\textrm{Rb},f_\textrm{Cs}) = (1,3)$ The results for rate coefficients dependent on $M_F$ at $E/k_\textrm{B}=20\ \mu$K are shown in table~\ref{table:rates-MF}. Contributions from s-wave and p-wave collisions are included. At sufficiently low field the rate coefficients are independent of the sign of $M_F$. It may be seen that atom pairs with different values of $M_F$ have significantly different rate coefficients. Spin-stretched atom pairs with $M_F=\pm5$ cannot decay collisionally at all, because there are no open channels. Even in the absence of Raman excitation, this would lead to non-exponential decay of the overall population.

The present experiments are not detailed enough to separate the decay for individual values of $M_F$. To allow comparison with the experiments, we carry out a further average over $M_F$ for each initial $(f_\textrm{Rb},f_\textrm{Cs})$, taking account of the remaining degeneracy. The resulting degeneracy-averaged rate coefficients for collision energy $E/k_\textrm{B}=20\ \mu$K are included in table~\ref{tab:k2s}. The calculations are approximately within the experimental error bars for $(f_\textrm{Rb},f_\textrm{Cs}) = (2,4)$ and (1,4), but rather outside them for (2,3). This can probably be attributed to the variation of rate coefficients with $M_F$ and to uncertainties in the distribution of the initial population among $m_{f,\textrm{Rb}}$ and $m_{f,\textrm{Cs}}$. Future experiments in a 1064~nm tweezer where spontaneous Raman scattering is suppressed should be able to resolve individual initial atomic pair states and allow more detailed comparisons with theory.



\section{Conclusions}

We have presented a detailed methodology for the efficient preparation of exactly one Rb and one Cs atom in the same optical tweezer. We have explained the considerations behind the choice of tweezer wavelengths and demonstrated that our tweezers are species-selective. We have demonstrated how to align and overlap the tweezers precisely in 3D, allowing the merging of tweezers containing single Rb and Cs atoms into the same optical tweezer by dynamically tuning the RF frequency applied to an AOD. This control was exploited to study 2-body collisions between Rb and Cs atoms in the pristine environment of an optical tweezer. We extracted hyperfine-state-dependent loss rate coefficients and compared them to theoretical results from coupled-channel calculations. Our results demonstrate that the precise control inherent in optical tweezer experiments holds great promise for the study of ultracold collisions of atoms and molecules.


The work presented in this paper represents a critical first step towards the creation of a single ground-state RbCs molecule. The next step will be to use Raman sideband cooling to prepare each species in the motional ground state of its respective tweezer~\cite{Kaufman2012, Sompet2017, Liu2019, Wang2019}, prior to the merging into a single tweezer. With the atom pair in the motional ground state of the tweezer, magnetoassociation will be used to form an RbCs Feshbach molecule~\cite{Takekoshi2012,Koppinger2014}. Finally, the molecule will be transferred to the rovibrational ground state using the established STIRAP transitions~\cite{Takekoshi2014, Molony2014, Molony2016}. The creation of single RbCs molecules in an array of optical tweezers will open up new opportunities for the exploration of molecular quantum gates \cite{Ni2018,Hughes2020} and encoding of qudits using multiple hyperfine  states of the molecule \cite{Sawant2020}. Applying the merging techniques developed in this work will also allow the study of atom-molecule and molecule-molecule collisions with precise control of the particle number~\cite{Cheuk2020}. Such an approach will provide valuable insight into the nature of molecular collisions~\cite{Mayle:2012,Mayle:2013,Gregory2019,Croft2020} and the loss of non-reactive species observed in experiments with bialkali molecules~\cite{Takekoshi2014,Park2015,Guo:2018,Christianen:2019,Gregory2019,Voges2020}.




%

\section{Acknowledgements}
The authors thank P.~D.~Gregory for early work on the experimental design and R.~Sawant for development of experimental simulations. The authors thank M. R. Tarbutt, J. Rodewald and J. Blunsden at Imperial College London for useful discussions on the design of the experiment. This work was supported by U.K. Engineering and Physical Sciences Research Council (EPSRC) Grant EP/P01058X/1 and Durham University. Data and analysis from this
work are available at doi:\href{http://doi.org/10.15128/r1k3569440m}{10.15128/r1k3569440m}.

\vspace{20pt}


\bibliographystyle{iopart-num}
\bibliography{references_2}

\end{document}